\documentclass[1p,authoryear]{elsarticle}
\usepackage{amsmath,amssymb}
\usepackage{graphicx}
\usepackage{subfigure}
\usepackage{defin}
\usepackage{color}
\usepackage{bm}

\def\sL{{\bf ---}}\def\dashL{{\bf $- -$}}\def\dotL{{\bf $\cdot\cdot\cdot$}}\def\dashdotL{{\bf $-\cdot-$}}
\def\GLML{\textcolor{blue}{\dashdotL}\,}\def\IML{\textcolor{black}{\dashL}\,}\def\PBLML{\textcolor{red}{\sL} }
\journal{Journal of Ship Research}

\begin{document}	
\begin{frontmatter}

\title{Physics-Based Learning Models for Ship Hydrodynamics}
\author[mit]{G.D.~Weymouth}
\ead{weymouth@mit.edu}
\author[mit]{Dick K.P.~Yue}
\ead{yue@mit.edu}
\address[mit]{Massachusetts Institute of Technology,
77 Massachusetts Avenue, Cambridge, 02139, USA}

\begin{abstract}
We present the concepts of physics-based learning models (PBLM) and their relevance and application to the field of ship hydrodynamics. The utility of physics-based learning is motivated by contrasting generic learning models (GLM) for regression predictions which do not presume any knowledge of the system other than the training data provided,
with methods such as semi-empirical models which incorporate physical-insights along with data-fitting. PBLM provides a framework wherein
intermediate models (IM), which capture (some) physical aspects of the problem, are incorporated into modern GLM tools to substantially improve the predictions of the latter, minimizing the reliance on costly experimental measurements or high-resolution high-fidelity numerical solutions. To illustrate the versatility and efficacy of PBLM, we present three wave-ship interaction problems: (i) at speed waterline profiles, (ii) ship motions in head seas, and (iii)  three-dimensional breaking bow waves. PBLM is shown to be robust and produce error rates at or below the uncertainty in the generated data, at a small fraction of the expense of high-resolution numerical predictions.
\end{abstract}


\end{frontmatter}

\section{Introduction}\label{sec: intro}

The state of the art in computational fluid dynamics (CFD) has advanced rapidly in the last decades, complementing advances in whole field quantitative experimental measurements. Advanced numerical methods and high performance computing capabilities have produced high fidelity simulations of ship hydrodynamics such as time domain predictions of the motions and flow around a fully appended ship, resolving details of the breaking bow and stern waves and the underlying turbulent flow \cite[e.g.][]{weymouth2007,Carrica2007}.
Such first-principle physics-based tools achieve accuracy through minimization of numerical and user errors. Despite increases in computing resources, these methods are still extremely computationally expensive (requiring hundreds of millions of grid points and hundreds of thousands of CPU hours), and generally still depend on highly educated and experienced users. As such, they are still mainly appropriate only for final stages of the design process.

Learning-based models offer an attractive alternative to first-principle physics-based simulations.  Such models can be extremely fast, with design evaluations often requiring less than a second on personal computers. In contrast to first-principle methods, learning models require an available set of example (or training) data, which we denote by: $\Datai$ for $i=1\ldots n$, where $y$ is the dependant variable to be estimated (e.g., drag, heave motion, etc) and $\x$ is the vector of independent variables (e.g., ship speed, incident wave frequency, ...). This data is typically obtained from experiments or high-fidelity simulations. The learning model seeks to approximate the unknown function $f$
from which the example data are generated: $y=f(\x)+\e$, where $\e$ is the random error in the measured/example data. The learning model $\Gen$ approximating $f$:
\be \Gen(\x|\c)\approx  f(\x) \ee
is `trained' by optimizing internal coefficients, $\c$, to minimize error on the training data. The predictive error of the model is then tested on points held out of the training set.

In this work, we will use the name generic learning model (GLM) for models $\Gen$ which approximate the unknown function $f$ using no system-specific information other than the training set.  There is a vast literature for GLMs, from basic least-squares polynomials fits to splines to non-linear multi-layer neural networks.
A common limitation of all GLMs is that they can only make accurate test predictions when available training data completely maps out the test space. Regions of the design space with few example data have a high risk of prediction errors, and filling in the gaps requires costly experiments or computations.
In addition, more complex physical systems typically necessitate more elaborate GLMs which must be trained with correspondingly larger sets of examples \citep{Evgeniou2000}. This can also lead to over-fitting; defined as when an overly flexible GLM is used to reduce the training error, only to find poor generalization to the unseen test data \citep{Vapnik1995,Burges1998,Girosi1995}. Utilizing techniques such as cross validation \citep{golub1979} and regularization can help minimize over-fitting (and will be exploited herein) but they do not address the systemic concern: a generic learning model does not know anything about the system other than the data it is provided.

This deficiency in GLMs is a well established historical problem and significant effort has been dedicated to the task of relieving the data dependence of such regression models through the addition of some level of `physical insight'. As an example, consider the scaling of the resistance of a ship proposed in \citet{Froude1955}. In modern terminology, Froude's insight was that a ship's resistance is a function of the Reynolds number $Re$ and the Froude number $Fr$ and that these factors act independently. In other words, the total resistance coefficient $C_T$ can be approximated by an additive model
$$ C_T = f(Re,Fr,\x_{geo}) \approx C_r(Fr,\x_{geo})+C_f(R_e) $$
where $C_f$, the friction coefficient is assumed to be a function only of $Re$, and $C_r$, the residuary resistance coefficient is a function only of $Fr$
(and the ship geometry represented by input geometry parameters  $\x_{geo}$). The decomposition of the $Re$ dependence is a physical insight which enables the friction model to be determined independently using flat plate data, thereby allowing data from $Fr$-scale model tests to be used in predictions of full-scale ship resistance.

The 1957 ITTC semi-empirical model for the friction coefficient $C_f$ is another example of a data-based model incorporating physical insights obtained from turbulent flat-plate boundary layer \citep{Granville1977}:
\be\label{eq: model friction}
    C_f(Re|\c) = \frac{c_1}{(\log Re -c_2)^2}.
\ee
In \eqref{eq: model friction}, $\c=\{c_1,c_2\}^T$ are the internal model coefficients, and setting $c_1\approx2$ and $c_2\approx0.075$ result in model predictions which agree with the experimental resistance data taken from a fully submerged flat plate to within 9\% over a wide range in the input ($Re=10^5\sim 10^{10}$). In comparison, a generic model function $\Gen$ in  polynomial form with the same number of free parameters
$$ \Gen(Re|\c) = c_1+c_2 Re $$
and a least-squares fit leaves regions of the data with more than 20\% error. The simple semi-empirical model \eqref{eq: model friction} illustrates how physics-based insights in a learning (regression) model can greatly increase predictive accuracy without increasing the number of free parameters.

Extending this general approach to predict the behavior of systems which are too complex to be modeled by an analytic semi-empirical model is an active research topic. Successful examples in the naval architecture literature are the ONR ship and submarine maneuvering prediction models of \cite{Hess2006a,Faller2005,Faller2006}. Unlike time-domain unsteady CFD ship motions predictions (such as \cite{Weymouth2005,Carrica2007}), these models feed pre-computed steady-state force predictions (along with other system variables such as relative speed and bearing) into a recursive neural network learning model to generate predictions of the resulting ship motions in real-time. Such research illustrates that a combination of a GLM (the recursive neural network) and a simplified physics-based model (the steady-state force predictions) with high-fidelity data can produce useful design tools. The ONR learning model has a complex structure which has been specially designed for maneuvering predictions.  The question we seek to answer is whether there is a general framework for such ideas that can be applied to a variety of ship hydrodynamics and ocean engineering problems, taking advantage of physics-based insights that might be obtained from preliminary design tools such as potential flow models.

In this work we develop a general approach for physics-based learning models (PBLM) which extends historical approaches such as analytic semi-empirical models and expands upon modern developments such as the ONR maneuvering predictions. The key idea behind PBLM is the robust incorporation of one or more physics-based intermediate models (IM) within a GLM.  Rather than increasing the complexity of $\Gen$ to improve the fit, PBLM complements a simple GLM with expanded basis functions constructed from fast and robust physics-based IM(s).  By complementing the GLM and the training data with physical behaviors (and constraints) inherent in the IM, we demonstrate that PBLM obtains significantly improved accuracy while reducing data dependence and the risk of over-fitting.

We develop our PBLM function $\Phy$ from three primary components: a training data set $\Data$; a generic learning model (GLM) defined by $\Gen$;  and a physics-based intermediate model (IM) defined by $\Intr$. (While the methodology easily generalizes to the use of more than one IM,  for clarity and simplicity we assume hereafter only one IM.).
The PBLM framework applies to many simple and robust GLMs, but to illustrate the power of PBLM, we  use regularized non-parametric GLMs, which are explained in \S\ref{ssec: GLM}, after covering some fundamental topics in machine learning. In \S\ref{ssec: PBLM}, our method for constructing $\Phy$ from these components is developed using a simple example of the transient development of a body in resonant motion.  To demonstrate the usefulness and performance of PBLM to realistic ship hydrodynamics applications, we apply this new  framework to three wave-ship interaction problems: (i) at speed waterline elevation profiles over a range of Froude numbers (\S\ref{ssec: waterline}), (ii) pitch and heave response of a ship with forward speed  (\S\ref{ssec: motions}), and (iii) breaking bow wave characteristic predictions for variable ship speed and breadth (\S\ref{ssec: tdpt}). The first two cases utilize the Wigley hull and are chosen because of their multi-dimensional input space and the wealth of classical experimental data for this canonical hull form. The last case introduces geometry dependence, non-linear wave mechanics, and uses computational sources for both the training data and the physics-based intermediate model. In all cases, we show that the PBLM approach achieves accuracy levels equivalent to high-fidelity experimental or CFD predictions, are significantly less data dependant than generic learning models, and have very small overall prediction costs.

There are many other branches of numerics which are related to the current work including the field of data assimilation using ensemble Kalman filtering \citep{Kalman1960,Evensen2003} and the field of multi-level optimization methods such as Approximation and Model Management Optimization (AMMO: \cite{Alexandrov1998,Alexandrov2001}). The ensemble Kalman Filter (enKF) is widely used in many fields such as climate and weather forecasting which rely on simplified models. AMMO is used to in many types of engineering optimization problems where evaluating approximate objective functions leads to great computational saving. 

The current work is influenced by these methods and others which use multiple levels of information to develop predictions, but the problem we have set out for ourselves is distinct. The the enKF is used to recursively filter data using large ensembles of simple forecasting models with randomized parameters. On the other hand, AMMO is utilized to establish trust regions for the approximate objective function to ensure convergence to the optimum of the true objective function. In contrast, the goal of the PBLM is to use an intermediate model with no free parameters to help generalize a small set of high fidelity examples to achieve accurate predictions across the input space. This could be seen as the sparse-data sparse-model limit of the filtering problem, or as an optimization problem which requires accurate predictions of the full response space. With these distinctions in mind, our development in the following section focuses on the literature of machine learning from labelled data as it is most directly applicable to our problem setting.

\section{Methodology}\label{sec: method}

To support the development of our PBLM approach, we first present some key concepts in machine learning. Consider a physical problem wherein we are given $n$ data points $\Datai$ for $i=1\ldots n$ generated by the unknown function $f$. The simplest generic model of this function would be a weighted sum of the inputs $\x$, ie
\be \label{eq:lin}\Gen(\x|\c) = \x^T\c \ee
which is linear in the inputs and the parameters $\c$. The model parameters are optimized to minimize an error metric $H(\Gen)$ such as the sum of squared error over the $n$ points in the training set:
$$H(\Gen) = || y_i-\Gen(\x_i) ||^2 = \sum_{i=1}^n \left[y_i-\Gen(\x_i)\right]^2 .$$

Unfortunately, a linear superposition of the input parameters \eqref{eq:lin} rarely provides an adequate fit to the data describing complex engineering systems. This can be improved by expanding the degrees of freedom using a parametric model
\begin{equation}\label{eq: parametric}
\Gen(\x|\c) = \g(\x)^T\c
\end{equation}
where $\g(\x)$ are a vector of generic basis functions, such as a polynomial basis $(1,x,x^2,\ldots)^T$. While the form of \eqref{eq: parametric} is simple, it still leaves the complex matter of defining the basis set $\g$. Using a low-dimensional basis with too few degrees of freedom will not help the fit, and least-squares fitting of high-dimensional basis are prone to over-fitting and can become numerically unstable.

Our main contribution to methodology is to introduce a physically motivated basis via a physics-based intermediate model (IM) to increase accuracy without over-fitting. This PBLM formulation is developed in \S\ref{ssec: PBLM}, but first we consider a generic learning model solution, non-parametric regularized learning.

\subsection{Generic learning models using non-parametric regularization}\label{ssec: GLM}

We introduce two generic machine learning approaches to help reduce training error without over-fitting; non-parametric regression and regularization. We discuss these methods because they introduce concepts such as effective degrees of freedom, and form the basis of the GLMs used in our PBLM approach.

Non-parametric regression methods such as splines and kernel interpolation ensure a good fit to the training data by generating a generic basis function for each data point. Typically non-parametric methods use a generic learning function of the form
\begin{equation}\label{eq: kernel sum}
	\Gen(\x|\c) = \sum_{i=1}^n c_i \Kern(\x-\x_i) =  \sum_{i=1}^n c_i \gi(\x)
\end{equation}
where $\Kern$ is a symmetric kernel function which is used to generate a generic basis functions $\gi$ for every data point $\x_i$. Unlike a polynomial basis, these kernels are localized, e.g.,  a Gaussian kernel $\Kern(\x)=\exp(||\x||^2/\sigma)$, limiting the influence of each basis function to a subregion of the input space. This enables the non-parametric model to accurately fit the training data from a wide variety of systems. However, non-parametric models have up to $n$ degrees of freedom which makes them very prone to over-fitting.

Regularization (also known as ridge regression) reduces over-fitting by introducing a `statistical-insight' into the GLM, namely: smooth predictions tend to generalize better to unseen data. To regularize a GLM, the parameters are optimized by minimizing a regularized error metric:
\begin{equation}\label{eq: regularized}
	H(\Gen) = (1-\lambda)||y_i-\Gen(\x_i)||^2 +\lambda J(\Gen),
\end{equation}
where the regularization function $J(\Gen)$ is a term which penalizes the higher derivatives of $\Gen$ and $\lambda$ is the regularization constant bounded by $0\leq\lambda<1$. The effect of regularized learning is a reduction in the effective degrees of freedom, eDOF, of the GLM, given by
\be\mbox{eDOF} = \sum_{i=1}^n\frac{(1-\lambda)d_i^2}{(1-\lambda)d_i^2+\lambda} \label{eq:edof}\ee
where $d_i$ are the eigenvalues of the linear system used to minimize Eq.~\ref{eq: regularized}.

The parameter $\lambda$ determines the trade-off between data-fit and model smoothness. Setting $\lambda=0$ results in an unconstrained least squares model with eDOF=$n$. This model would be flexible, but would over-fit the $n$ training data points. When $\lambda\rightarrow1$, the function $\Gen$ is constrained to be linear, with eDOF$\rightarrow0$. This model would be robust, but is not likely to predict the system behaviour accurately. An intermediate value of $\lambda$ is determined automatically as in \cite{golub1979} by minimizing the generalized cross validation error on the training set in an attempt to balance the desire for high eDOF to fit the training data and the desire for low eDOF to smooth $\Gen$ and improve generalization.

We use three well-established versions of regularized non-parametric regression in our applications below: thin plate regression splines \citep{Wood2003,Wood2008}, gaussian regularization networks \citep{Girosi1995,Burges1998}, and support vector machines \citep{Vapnik1995,Evgeniou2000,Scholkopf2000}. In all cases, the regularized problem is more numerically stable than least-squares and has a unique solution for the optimum parameters, unlike non-linear methods such as neural networks. However, the model performance is still dependant on the amount of training data and is highly sensitive to the chosen value of $\lambda$ (and any parameters in the kernel function $\Kern$).

\subsection{Physics-Based Learning Models}\label{ssec: PBLM}

The regularized non-parametric GLMs presented in the preceding section are flexible and offer a level of protection from over-fitting but they are still generic methods and cannot make reliable predictions in data-sparse regions of the input space. We will reduce this data dependence not by fine-tuning parameters or adopting more complex GLMs but by enhancing these robust learning tools with physics-based information from an intermediate model $\Intr$. Additionally, instead of trying to develop a new learning approach for each new system, we develop a general PBLM approach wherein the physics-based learning model $\Phy$ is of the form
\be\Phy(\x) = \Gen(\x,\p(\x)),\ee
augmenting a simple GLM with a physics-based basis $\p$ to reduce data-dependence and increase accuracy.

\begin{figure}
    \centering
    \subfigure[Data-based GLM]{
    	  \label{fig: toy a}
          \includegraphics[height=.35\textwidth, clip, trim=10 10 20 20]{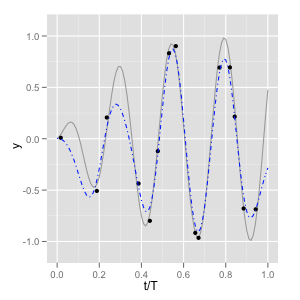}}
    \subfigure[IM and additive model]{
    	  \label{fig: toy a2}
          \includegraphics[height=.35\textwidth, clip, trim=47 10 20 20]{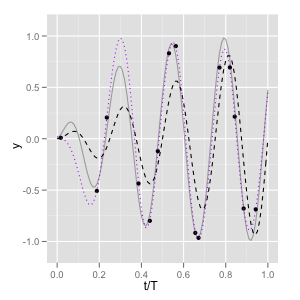}}
    \subfigure[PBLM]{
    	  \label{fig: toy b}
          \includegraphics[height=.35\textwidth, clip, trim=47 10 20 20]{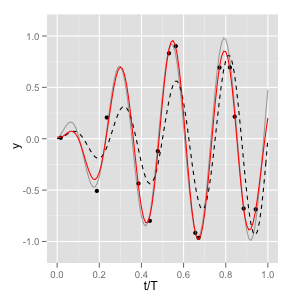}}
    \caption{PBLM example for the transient startup of harmonic motion.  In all figures \textcolor[gray]{0.5}{\sL} is the true system function $f$ and $\bullet$ are the $n$=15 randomly sampled values of $f$ with Gaussian noise of $\sigma$=10\%. In figure (a), \GLML is the optimized Gaussian regularization network GLM fit. In figure (b), the true function $f$ is compared to \IML the simple IM $\Intr$ \eqref{eq:toyIM}, and \textcolor{magenta}{\dotL} the predictions of the additive model $f_{\cal A}$ \eqref{eq: line add}.  In figure (c) \PBLML is the prediction of PBLM \eqref{eq: phase basis} using the phase-shifted basis $\p$.  The PBLM results are more than twice as accurate than either the GLM or the additive model, despite the small training sample and the fact that the IM has 32\% error.
    }
    \label{fig: toy}
\end{figure}

To illustrate the basic ideas and efficacy of PBLM, consider a simple model problem of the transient startup from rest of the resonant harmonic motions, say corresponding to initial development of the resonant response of a floating body in regular waves.
For simplicity, we assume the true motion is given by
\be f(t) = \alpha \tanh(\beta t) \sin(\omega t +\phi) \label{eq:true}\ee
where $\beta$ is the startup time scale, $\omega$ is the frequency of the regular incident wave, and $\phi$  phase difference between the response and the wave; and that  we have a limited sample of experimental data of this system
\be y_i = f(t_i)+{\cal N}(\sigma),\quad i=1\ldots n, \label{eq:datan}\ee
at $n$ randomly sampled times, $t_i$, where the $\cal N$ term is a zero-mean Gaussian noise in the data (see Fig.~\ref{fig: toy}).

The first approach is to fit \eqref{eq:datan} with a GLM. In this case, linear and parametric GLMs give universally inadequate predictions because of the high frequency and variance in the system response. Non-parametric GLM results are highly dependant on the form of the kernel $\Kern$, the regularization parameter $\lambda$, and the exact data sample provided. Fig.~\ref{fig: toy a} shows the predictions of an optimized Gaussian regularization network GLM. The result are representative of a well-tuned model, with the fit being better in regions with more data and worse in region with less, with an overall RMS error of ~20\%. The predictions, however, are not robust, with high sensitivity to changes in the training data and parameter values.

To improve the performance of the GLM for this sparsely sampled system, we introduce a simple IM which captures merely the physics-based expectation that (a) in steady state, the response should be harmonic with frequency $\omega$; and (b) the initial resonant response grows linearly in time.  Without additional knowledge of the motion amplitude or phase, or the time required to reach steady state, we construct an IM simply as, say,
\be\Intr(t) = t \sin(\omega t). \label{eq:toyIM}\ee
This function, after appropriate scaling, is plotted in Fig.~\ref{fig: toy a2} and while qualitative similarity to the true response, it has a quantitative root-mean square error (RMSE) of over 30\%.

While \eqref{eq:toyIM} has low predictive accuracy, it encapsulates additional knowledge of our system and can be used to improve the GLM prediction by supplementing the data with this information. In this simple example, we could parametrize $\Intr$, turning it into a semi-empirical model. However, this approach does not apply to general intermediate models. The simplest way to use the IM as a `black-box' is to treat the difference between $\Intr$ and the data as a nuisance variable and fit that error with a GLM. Taking this idea one step further, we can use a linear additive model \citep{Scholkopf2001,Wood2008}
\begin{equation}\label{eq: line add}
	f_{\cal A}(t|\c) = \Intr(t) c_\rho + \Gen(t|\c_\gamma)
\end{equation}
which linearly weights the IM predictions with GLM predictions to obtain the best overall fit to the data. The additive model is an improvement over the nuisance model because the physics-based coefficient $c_\rho$ is determined simultaneously with the generic coefficients $\c_\gamma$. The nuisance and additive models incorporate information from intermediate models in a general way, but they will only improve upon a GLM when the IM is a very good predictor of the system. Fig.~\ref{fig: toy a2} shows that the additive model predictions for this  example system are only marginally better than the GLM results, with the RMSE near 16\%. Additionally, the performance is still highly dependant on the details of the GLM and the exact data sampled.

While the additive model is a step in the right direction, it is too reliant on the GLM and the accuracy of the function $\Intr$. Our solution is to expand the physics-based degrees of freedom into a representative basis using a Taylor expansion of $\Intr$ (to provide gradient information). This gives the learning model more information about the form of $\Intr$ and more flexibility in constructing a physics-based solution before relying on the generic basis $\g$. Development and evaluation of an exact Taylor expansion for general IMs would be a significant expense. However, one of the central advantages of shifting our information burden from the data to the IM is that the IM can be evaluated anywhere very quickly, not just at the x points where the data is available. Therefore, we can establish properties of the IM, such as the local derivatives, using a finite difference approximation
$$ \p(x) = \left(\Intr(x), \frac{\Intr(x+\D x)-\Intr(x-\D x)}{2\D x},
\frac{\Intr(x+\D x)-2\Intr(x)+\Intr(x-\D x)}{\D x^2}\ldots\right)^T.$$
where the points $x\pm \D x$ are not required to be in the set of points $x$ for which we have training data. Through linear combinations, the finite difference is equivalent to
\begin{equation}\label{eq: phase basis}
\p(\x) = (\Intr(\x),\Intr(\x+\D \x),\Intr(\x-\D \x))^T.
\end{equation}
This phase-shifted physical basis is equivalent to the derivative basis in the limit of $\D x\rightarrow0$, is easily computable using the intermediate model $\Intr$, and enables the generation of robust and accurate PBLMs. In cases where phase-shifting either left or right of a data point moves past a physical boundary (such as zero speed) the offending term is simply left out and the implied derivative reverts to a one-sided first-order approximation. 

We develop two simple and robust methods of including the phase-shifted physical basis $\p$ in our PBLM. First, we can construct a parametric additive model
\begin{equation}\label{eq: additive}
	\Phy(\x|\c) = \p(\x)^T\c_\rho+\Gen(\x|\c_\gamma)
\end{equation}
which complements the generic basis $\g$ with the physical basis $\p$. A non-parametric version of the form of \eqref{eq: kernel sum} can also be constructed using $\p$ to complement the input space, i.e.,
\begin{equation}\label{eq: weighted}
\Phy(\x|\c) = \Gen(\z|\c) = \sum_{i=1}^n c_i \Kern(\p(\x)-\p(\x_i))\Kern(\x-\x_i)
\end{equation}
where $\z \equiv (\x^T,\p(\x)^T)^T$. This weighted kernel formulation introduces physical-relevance into the PBLM while utilizing the advantages of localized non-parametric regression. The predictions using Eq.~\ref{eq: weighted} are shown in Fig.~\ref{fig: toy b}. The accuracy is excellent, with an RMSE of 8\%, lower than the noise in the data and less than half of the GLM error. Additionally, this performance level is robust to changes in the details of the GLM and the training sample. 

This simple one dimensional example turns out to be quite representative of the later results for realistic complex problems. Furthermore, our PBLM construction above is completely general, requiring no adjustment or modifications to produce predictions for real ocean engineering systems with multidimensional input spaces, as illustrated in the later applications in \S\ref{sec: applications}.

We have developed both \eqref{eq: additive} and \eqref{eq: weighted} to demonstrate that physics-based learning is a general concept which may be used in different specific machine learning models. The parametric form of \eqref{eq: additive} is simpler, and more familiar to anyone who has used a linear solver. The non-parametric form of \eqref{eq: weighted} is more flexible, but it requires a bit of additional machinery to implement. Additionally, different types of learning problems may dictate different model structures. Parametric models can be more readily interpreted by examining how the basis functions have been weighted, whereas problems with Runge's phenomenon may be avoided using a non-parametric approach. We will use \eqref{eq: additive} in our first example in \S\ref{ssec: waterline} and \eqref{eq: weighted}  is used in our examples in \S\ref{ssec: motions} and \ref{ssec: tdpt}.

Before moving to the applications, we remark on the desired attributes of the IM used to construct the physical basis $\p$. As in the example above, the ideal $\Intr$ should be efficient, stable, robust, provide good descriptions or approximations of (some aspects of) the system,  and does not itself depend on free parameters or require new system-specific learning or user artistry. Note that the inclusion of an intermediate model function $\Intr$ which is a poor approximation of the system, while not ideal, is also not catastrophic. An inaccurate IM will be poorly correlated to the training data, and will be ignored by the learning machine (given no weighting) meaning the PBLM will simply preform as a GLM. An illustration of this is shown in the example in \S 3.2. A more serious constraint is that the IM prediction be noise free. As the IM is used to construct a basis, any noise in $\Intr$ would propagate up to $\Phy$ giving undesirable noisy final predictions. Practically, this can almost always be achieved by filtering away background noise in the IMs before using them in a PBLM as demonstrated in the last example problem of \S 3.3.

\section{Applications}\label{sec: applications}

To illustrate the application and performance of the PBLM method developed above, we consider in this section three ship hydrodynamics applications involving the prediction of (i) the waterline profiles on a ship over a range of Froude numbers, (ii) the heave and pitch responses of a ship advancing in regular seas, and (iii) the breaking bow waves for a ship of varying forward speed and beam-to-length ratio.

\subsection{Waterline Predictions for the Wigley Hull}\label{ssec: waterline}

We consider the predictions of the waterline on a  Wigley hull over a broad  range of Froude numbers.
These classical experiments have served as benchmark data for many numerical methods including advanced potential flow and Navier-Stokes simulations. Even for this case of steady flow and relatively smooth profiles, a generic learning model trained using experimental data set has large errors in data-sparse regions and makes non-physical predictions. We demonstrate that by complementing the GLM with a simple linear potential flow intermediate model, PBLM  achieves excellent predictive accuracy across the input space, even at untrained Froude numbers.

\begin{figure}
    \centering
    \includegraphics[width=\textwidth, clip, trim= 10 10 20 10]{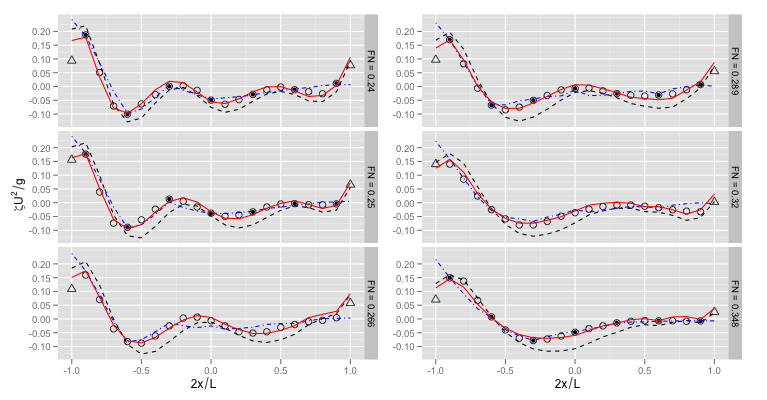}
    \caption{ Waterline elevation profile data and predictions for the Wigley hull over six Froude numbers. Symbols denote the experimental measurements: $\bullet$ are the training data, $\circ$  are the test data, and $\triangle$ are  extrapolation points outside the training data. Lines denote the prediction methods: \IML is the potential flow IM prediction, \GLML is the GLM prediction (using thin plate regression spline), and \PBLML the PBLM prediction.  PBLM obtains significantly increased accuracy away from training data, interpolating accurately to the unseen Froude numbers, and extrapolating to recover physically reasonable predictions for the bow and stern elevations.}
    \label{fig: waterline}
\end{figure}

The data used for training and testing our models comes from the experimental measurements of \cite{Shearer1965} which reports the at-speed waterline profile for a fixed sinkage and trim Wigley hull. From this source we select data for the normalized waterline elevation $\eta U^2/g$ over six $Fr$ speeds and 21 $2x/L$ section positions. These waterline profile measurement points are plotted  in Fig.~\ref{fig: waterline},  showing the dependence of the profiles on Froude number.

For the GLM, we use thin plate regression splines of \cite{Wood2003}, a regularized multi-dimensional spline method. We train this generic learning model to predict the waterline profile over the two-dimensional input vector by minimizing the regularized error (Eq.~\ref{eq: regularized}) over the training set $\{y=\eta U^2/g\;|\,\x=(Fr,2x/L)^T\}$. A third of the profile points were included in the training data except for two intermediate Froude numbers ($Fr$=0.266, 0.32 in Fig.~\ref{fig: waterline}) which were completely held out to test the model's ability to generalize to unseen speeds. The held out points within the convex hull of the training data make up the test set, used to determine the generalized accuracy of the GLM predictions. Predictions of the elevation near the bow and stern are extrapolations outside the training data, making these especially difficult for the GLM. With the model trained, the GLM predicts the wave elevation at every point in the parameter space using  trivial computational effort.  Considering the sparsity of the supplied training data set, the thin plate regression spline GLM performs fairly well as summarized in Tab.~\ref{tab: waterline}. The test RMSE is around 20\% of the maximum waterline elevation measurement. However, as expected of any generic learning model, the error increases for predictions far from the training data, for example in the unseen Froude numbers. The extrapolated predictions for the leading and trailing edge wave heights are also non-physical, missing the low initial wave height on the bow and failing to predict the run up at the stern.

For the intermediate model we use a first-order potential flow panel method and a reflection (double-hull) boundary condition for the free surface as in the classical approach of \cite{Hess1964}. This simple model is chosen for the speed and robustness of the IM predictions, rather than their accuracy. All of the potential flow numerical parameters (related to the panelization, inversion etc.) are fixed to avoid propagation of learning capacity to the PBLM. The waterline predictions using this IM are plotted in in Fig.~\ref{fig: waterline}. As summarized in Tab.~\ref{tab: waterline}, these predictions have an RMSE similar to the GLM, around 20\%.

\begin{table}
  \centering
  \begin{tabular}{rccc}
   			              & IM (potential flow)& GLM & PBLM  \\[3pt]
	Test set RMSE & 0.193 & 0.201 & 0.085 \\
    Total  RMSE & 0.187 & 0.177 & 0.078 \\
                eDOF     &   0      & 21.2   &   5.0 \\
  \end{tabular}
  \caption{RMSE and effective degrees of freedom (eDOF, Eq.\eqref{eq:edof}) for the prediction of the Wigley hull waterline elevation.   The RMSE, scaled by the maximum response $\eta U^2/g=0.187$, are given for the test set as well as for the total (including the error on the training, test, and extrapolation points) compared to tank measurements.  The GLM is the thin plate regression spline, and the IM is simple potential flow with reflection free-surface boundary condition. The RMSE of the IM and GLM are comparable (close to 20\%), while the PBLM incorporating these (using the phased-shifted physics-based basis with only 5 effective degrees of freedom) is able to reduce the test and total errors but a significant factor.}
  \label{tab: waterline}
\end{table}

For the PBLM, we construct the phase shifted basis $\p$ of Eq.~\ref{eq: phase basis} using the potential flow IM. The basis is incorporated into the GLM through the simple additive model of Eq.~\ref{eq: additive} and retrained on the training data. The predictions for the PBLM are shown as the red solid line in Fig.~\ref{fig: waterline} and summarized in Tab.~\ref{tab: waterline}. The addition of $\p$ has reduced the error on the test data by 58\% and the figure shows the accuracy is much less dependent on the proximity of the training data. The figure also shows that the predictions of the trailing edge height are excellent despite these points being extrapolations beyond the training data. The leading edge profile heights are generally over-predicted, but always physically realistic.

The effective degrees of freedom (eDOF) of GLM versus PBLM are given in Tab.~\ref{tab: waterline} quantifying the decreased data dependence of the PBLM. The potential flow model has no adjustable degrees of freedom (eDOF=$0$) and the GLM has effectively 21 degrees of freedom after regularization. As there are only 28 points in the training set, the GLM is only weakly regularized, which translates to higher data dependence. In contrast, PBLM has very little data dependence, using only 5 degrees of freedom to fit the data more than twice as accurately. 

\begin{figure}
    \centering
    \subfigure[]{
          \includegraphics[width=0.45\textwidth]{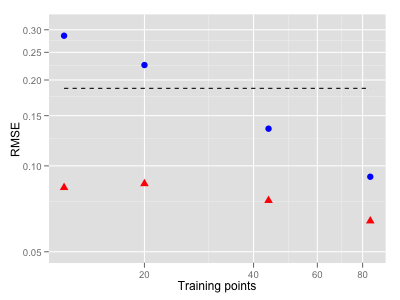}}
    \subfigure[]{
          \includegraphics[width=0.45\textwidth]{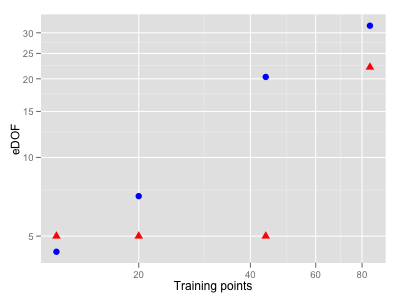}}
  \caption{Convergence study on the dependence of the test RMSE (a) and eDOF (b) on the number of training points for the GLM (\textcolor{blue}{$\bullet$}) and PBLM (\textcolor{red}{$\blacktriangle$}). The RMSE is scaled by the maximum response $\eta U^2/g=0.187$ and the potential flow RMSE is shown for reference (\IML). The training sets were selected using systematic reduction, and in all cases the $Fr$=0.266,0.32 data have been held out. Note that the PBLM is essentially converged using 12 training data points (3 points per speed), and at this level outperforms the GLM using 84 points.}
  \label{fig: waterline study}
\end{figure}

A systematic study further quantifies the dependence on the number of training points and is presented in Fig.~\ref{fig: waterline study}. For this study, the number of waterline points used in the training set was systematically reduced using a `point-skipping' approach; starting with all 21 point in each profile, then using every other point, and eventually ending up with only the points at $2x/L=-1,0,1$ in the training data. The points in each coarse set are contained in all the finer sets, as in a nested quadrature.
In all cases the $Fr$=0.266,0.32 data have been held out of the training set. The results in Fig.~\ref{fig: waterline study} show that the GLM response is completely dependent on the number of training points, making steady gains in accuracy with increased data (and increased eDOF). In contrast, with as few as 12 training points the PBLM solution has essentially converged on the solution shown in Fig.~\ref{fig: waterline} with RMSE$\approx$8.5\%, which is more accurate than the GLM solution using all 84 available waterline points. Only with the full set of points does the PBLM relax it's regularization and increase from 5 to 22 degrees of freedom to more closely fit the data, achieving 6.4\% test RMSE. These results quantify that the addition of the physics-based basis $\p$ is remarkably effective at shifting the learning dependence from the expensive data measurements to the inexpensive intermediate model. 

\subsection{Pitch and Heave Motions Predictions}\label{ssec: motions}

We next consider the prediction of the heave and pitch response amplitude operators (RAO) for the Wigley hull  in head seas. As in the previous example, this problem has been used for the validation of many numerical prediction methods including the unsteady-RANS CFD predictions of \cite{weymouth2007}. The RANS predictions were found to be quite accurate with RMSE levels of around 2.5\%, but each $(Fr,\lambda/L)$ evaluation took $O(10^4)$ CPU hours.
We show that by incorporating a simple linear potential flow IM into a standard GLM we achieve a similar level of accuracy over the complete input space as \cite{weymouth2007} using a fraction of a second of computing time.  We also find that the PBLM predictions are robust to changes in the training data.

\begin{figure}
    \centering
    \subfigure[Heave]{
          \includegraphics[width=\textwidth, clip, trim= 10 10 20 10]{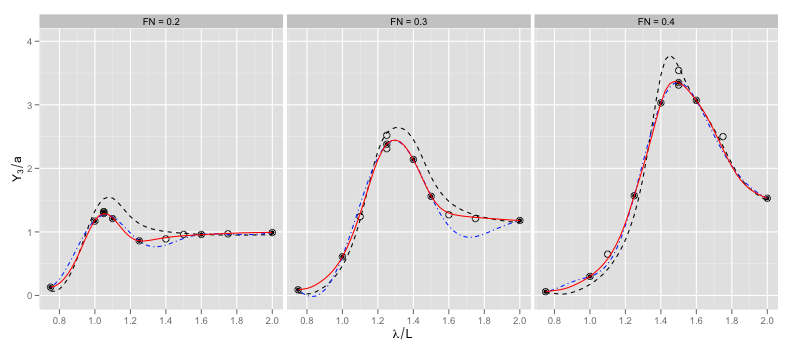}}
    \subfigure[Pitch]{
          \includegraphics[width=\textwidth, clip, trim= 10 10 20 10]{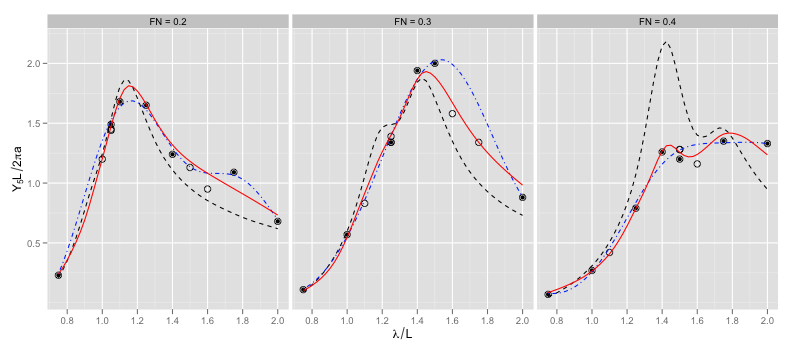}}
    \caption{ Heave and pitch response amplitude operators for the Wigley hull over a range of incident wavelengths and ship Froude numbers. Symbols denote the experimental measurements: $\bullet$ are the training data, $\circ$ are the test data. Line denote the prediction methods being compared: \GLML is the GLM (using Gaussian regularization network); \IML is the IM (using LAMP-I); \PBLML is the PBLM incorporating the IM in the GLM. The PBLM shows significantly increased accuracy away from training data, and correctly ignores the inaccurate LAMP-I pitch predictions for $Fr=0.4$.}
    \label{fig: heave1}
\end{figure}

The data for this problem is taken from the experimental measurements of \cite{Journee1992} for the Wigley hull I and wave amplitude $a/L\approx 0.005$. For this example we use data for the normalized heave ($Y_3/a$) and pitch ($Y_5 L/2\pi a$) response over three $Fr$ speeds and $\sim$11 $\lambda/L$ head sea wavelengths. The data is denoted by the point symbols in Fig.~\ref{fig: heave1} showing substantial  variations in the RAO peak  frequency and height with changing Froude number, reflecting the complex wave-ship-flow  interactions at play.

For this  example,  we choose the Gaussian regularization network GLM which is studied extensively in machine learning \citep[e.g.][]{Evgeniou2000}. This GLM  is trained on the training set, and compared against the test data in Fig.~\ref{fig: heave1} and quantified in Table~\ref{tab: heave1}. The GLM predictions are generally good, but as before, the performance deteriorates in data sparse regions, for example near $\lambda/L=1.5\sim 2.0$ for $Fr$=0.3.

For the physics-based IM we use the linear version of the Large Amplitude Motion Prediction (LAMP-I) program \citep{Lin1990}, a time-domain three-dimensional potential flow panel method.  The choice of LAMP-I as IM is based on the speed and stability of the linear potential-flow code. As in the previous application, all the model/computational parameters in LAMP-I are fixed, and the resulting predictions are shown in Fig.~\ref{fig: heave1}. As seen in the figure and in Table~\ref{tab: heave1}, the LAMP-I heave predictions are fairly accurate with RMSE=5.4\% of the maximum response amplitude. However, the pitch accuracy is lower, particularly for the high speed case.

\begin{table}
  \centering
  \begin{tabular}{rccc}
Heave  & IM (LAMP-I) & GLM & PBLM  \\[3pt]
Test set RMSE & 0.047 & 0.042 & 0.025 \\
Total  RMSE & 0.054 & 0.027 & 0.016 \\
eDOF & 0 & 16.4 &  10.3 \\
\\
Pitch  &IM(LAMP-I) & GLM & PBLM\\
Test set RMSE & 0.131 & 0.079 & 0.029 \\
Total  RMSE  & 0.141 & 0.053 & 0.030 \\
eDOF & 0 & 13.0 & 7.8 \\
  \end{tabular}
  \caption{Prediction errors for the test set and the total RMSE (which includes the error on the test and training points) and the effective degrees of freedom (eDOF) for the Wigley hull pitch and heave tests. The root mean square errors are scaled by the maximum response for each test ($Y_3/a =3.54$, $Y_5 L/2\pi a=2.0$). The PBLM demonstrates error levels below 3\% (equivalent to the error level of time domain Navier-Stokes simulations) using simple linear models and extremely sparse training data.}
  \label{tab: heave1}
\end{table}

For the PBLM, we construct the phase shifted basis $\p$ of Eq.~\ref{eq: phase basis} using LAMP-I as the IM. The basis is incorporated into the GLM through the weighted model of Eq.~\ref{eq: weighted} and retrained on the training data. The PBLM results are shown in Fig.~\ref{fig: heave1} and the RMSE and eDOF summarized in Table~\ref{tab: heave1}. PBLM using a simple linear IM and 20 training points produce predictions with error levels below 3\%, matching the performance of the time-domain RANS predictions of \cite{weymouth2007}. Moreover, the PBLM achieves this using only 10 degrees of freedom as opposed to the 16 degrees of freedom used by the GLM, whose predictions have around twice the error. Note that the inaccuracy of the IM in the high speed pitch case is not propagated to the PBLM response. As discussed in the introduction, an inaccurate IM is simply ignored by the learning model and the resulted PBLM error matches that of the GLM in this region. Of more concern is the large variance in the IM for this test case which has caused some waviness in the PBLM. Low-pass filtering the IM prediction alleviates the waviness in the PBLM prediction but does not significantly change the error levels.

\begin{figure}
    \centering
    \subfigure[No peak training data]{
          \includegraphics[width=\textwidth, clip, trim= 10 10 20 10]{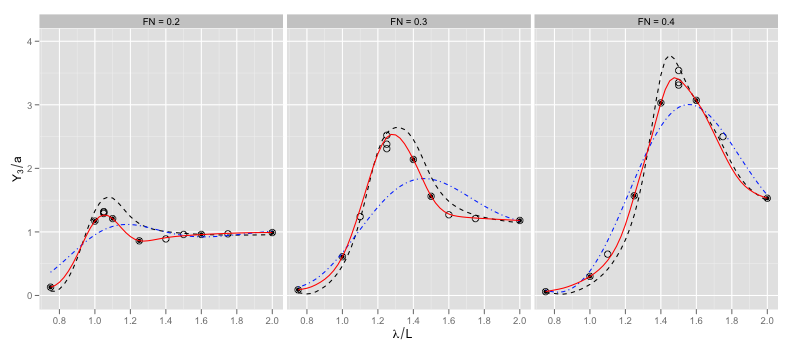}}
    \subfigure[No $Fr=0.3$ training data]{
          \includegraphics[width=\textwidth, clip, trim= 10 10 20 10]{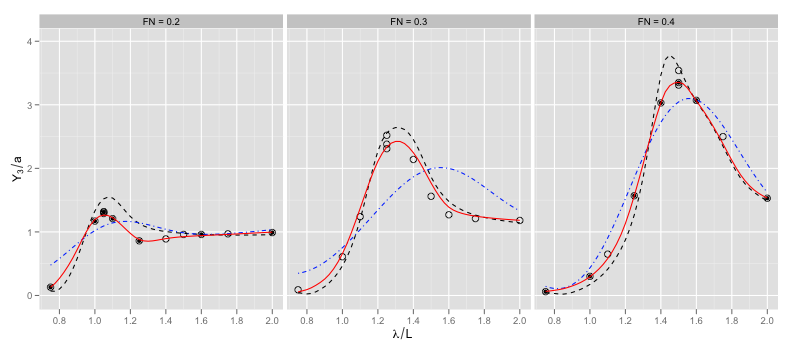}}
    \caption{ Heave RAO predictions for the Wigley hull using a reduced training set.  The symbols and lines designations are as in Fig.~\ref{fig: heave1}. Relative to Fig.~\ref{fig: heave1}, the training set is reduced by (a) removing the peak response points; and  (b) omitting all  $Fr=0.3$ data. In both cases, the PBLM prediction maintains its accuracy despite the removal of data; while the GLM performance is significantly affected.}
    \label{fig: heave2}
\end{figure}

To explicitly demonstrate the data-dependence of the GLM predictions and the improvement in the PBLM approach, we repeat the heave predictions with two modified training sets.  We note that our training set for the heave case is nearly ideal, spanning all three Froude numbers and includes all of the response peaks. While this yields accurate predictions, the situation is somewhat artificial.
To evaluate the robustness of the different predictions with more limited training data, we show in Fig.~\ref{fig: heave2} and Tab.~\ref{tab: heave2} results for the same problem but now withholding from the training set: (a) peak response values; and (b) data for the intermediate Froude number. In case (a), seemingly crucial data is omitted; while in case (b), the total number of training points is reduced to only 14.  In both cases, the PBLM performance is nearly unchanged with RMSE below 3\%. In contrast, the GLM performance is substantially degraded with more than three times the error.  It is possible that a more specialized GLM could be more appropriate for these extremely sparse data sets. However, the introduction of the IM and the physics-based basis $\p$ has eliminated the need for a more advanced learning method in these data-sparse cases.

\begin{table}
  \centering
  \begin{tabular}{rccc}
 & Training points & GLM & PBLM  \\[3pt]
Ideal Training Set & 20 & 0.042 & 0.025 \\
No peak data & 17 & 0.128 &  0.028 \\
No $Fr=0.3$ data & 14 & 0.131 & 0.027\\
  \end{tabular}
  \caption{Dependence of the Test RMSE on the training data set for the Wigley hull heave RAO predictions. The PBLM demonstrates a robust error level, $<$ 3\% despite the removal of critical training data.}
  \label{tab: heave2}
\end{table}

\subsection{Predictions of Breaking Bow Waves}\label{ssec: tdpt}

Finally, we consider the prediction of the nonlinear bow waves of a theoretical hull form based on the 5415 with a sharp bow. This example is motivated by the desire to apply fast 2D+T modeling to the prediction of breaking bow waves and to demonstrate the ability of PBLM methods on a design-stage problem with no available experimental data.

\begin{figure}
    \centering
    \subfigure[Test geometry station lines]{
	    \label{fig: 3dship a}
    	\includegraphics[height=0.35\textwidth, clip, trim= 10 0 40 10]{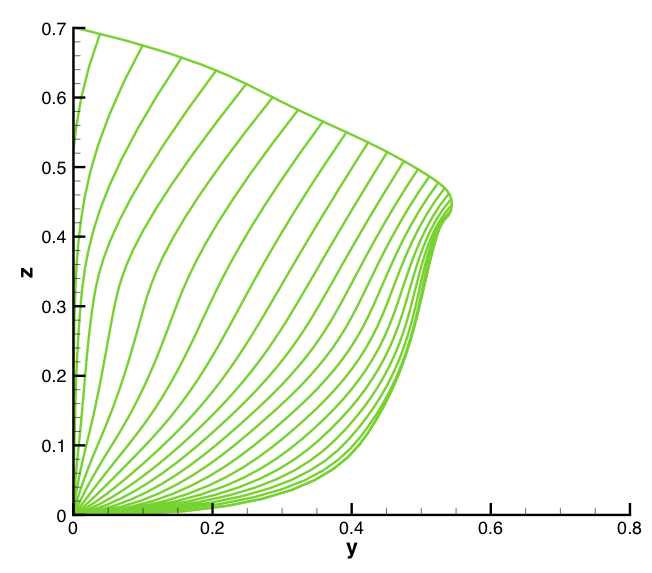}}
    \subfigure[3D simulation $F_{2D}=0.262$,$B/L=0.12$ ]{
    	\label{fig: 3dship b}
    	\includegraphics[height=0.35\textwidth, clip, trim= 0 -40 0 0]{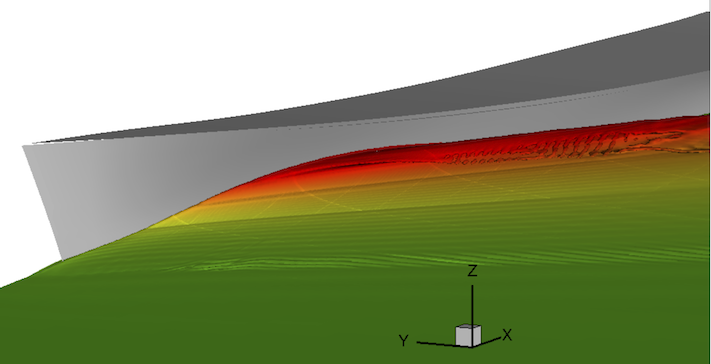}}
    \caption{Geometry and sample 3D simulation for the bow wave test case. The geometry section lines (a) are based on a 5415 hull modified to have a very fine entrance angle. The full 3D CFD simulation (b) shows the hull in grey and the free surface colored by elevation. For this case ($F_{2D}=0.262$,$B/L=0.12$) there is a large plunging breaker at the bow.}
    \label{fig: 3dship}
\end{figure}

\begin{figure}
	\centering
	\includegraphics[width=\textwidth, clip, trim= 10 10 20 10]{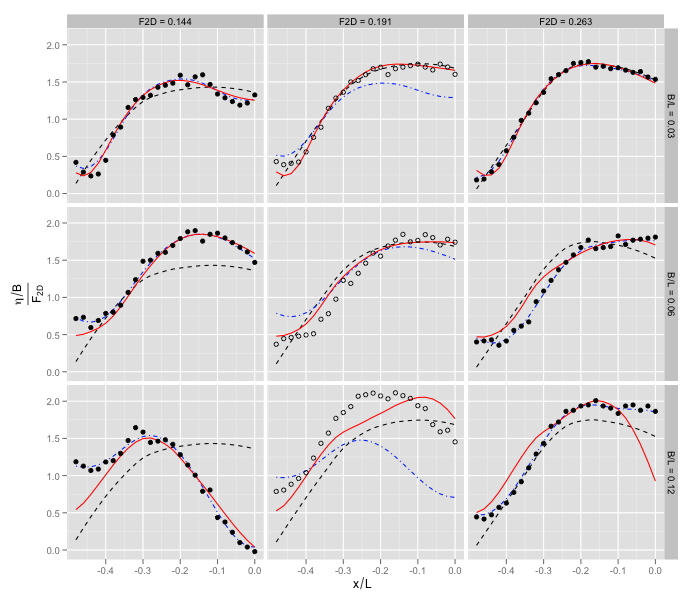}
	\caption{Data and predictions for the wave crest elevation of breaking bow waves at a function of 2D Froude number ($F_{2D}$) and ship thickness ($B/L$). Symbols denote measured values from the 3D Cartesian-grid CFD:  $\bullet$ are the training data, $\circ$  are the test data. Lines denote the prediction methods: \IML is the 2D+T IM prediction; \GLML the GLM prediction;  and \PBLML the PBLM prediction. PBLM enables predictions across the Froude number range with higher accuracy than the GLM even for larger $B/L$ values.} \label{fig: tdpt}
	\label{fig: crest}
\end{figure}

The data for training and testing the learning models is obtained in this case from full-3D high-fidelity CFD simulations using the cartesian-grid Boundary Data Immersion Method \citep{weymouth2011b} coupled with a conservative Volume of Fluid method \citep{Weymouth2010}. Simulations are run over three $B/L$ slenderness values and three $F_{2D}= Fr \sqrt{\frac B L}$ speeds. This scaling ensures that the speeds span the range of different types of bow wave generation for each slenderness value. The low speed gives of a non-breaking bow wave, the intermediate speed gives a gentle spilling breaker, and the high speed produces a plunging breaker. As an example, the ship hull and the 3D CFD simulation prediction for the high speed, high thickness result is shown in Fig.~\ref{fig: 3dship}. The simulations are processed to extract the height of the wave crest as a function of relative position down the length of the bow ($x/L$). The data is shown as the symbols in Fig.~\ref{fig: crest} where it can be seen that the extraction function introduces a level of background noise.

For the GLM in this case, we use the Support Vector Machine Regression \citep{Vapnik1995}. The training data supplied to the GLM include the high and low speed data, with the intermediate speed data withheld for testing. The GLM predictions are shown in Fig.~\ref{fig: crest} and summarized by $B/L$-value in Tab.~\ref{tab: crest}. As with the other GLMs, Support Vector Machine Regression shows good ability to fit the training data and correctly ignores the noise. However its generalization to the unseen Froude number test data is fairly poor, especially for the largest $B/L$. In order for a generic learning model to make accurate predictions for this case, many more 3D simulations would need to be run across the range of $F_{2D}$, at a significant increase of overall computational cost.

For the physics-based IM, we use a Cartesian-grid 2D+T model. The 2D+T approach is  formally based on slender body theory assuming that changes in the longitudinal direction are small compared to changes in the transverse directions \cite[e.g.][]{Fontaine1997,Fontaine2001,Weymouth2006}. In this approach, the 2D+T representation of the ship geometry is that of a 2D flexible wavemaker, whose instantaneous profile matches the section lines of the 3D hull. The successive 2D waves generated by this wave-maker correspond to the divergent waves of the 3D ship in the limit of high ship speed and slenderness. As there is no longitudinal length or velocity in the wavemaker flow, the 2D+T `speed' is set through the time $T=L/U$ it takes for the wavemaker to trace the  sectional shape of the vessel. Thus the appropriate non-dimensional scaling is $ F_{2D} = \sqrt{\frac B{gT^2}} $. Although theoretically only valid for a slender ship moving at high speed, approximating the actual three-dimensional flow by a two-dimensional, time evolving flow reduces the computational cost by orders of magnitude, making it attractive for use as the IM in a PBLM.

\begin{figure}
	\centering
    \subfigure[2D+T waterfall plot]{
	    \label{fig: tdptcrest a}
    	\includegraphics[width=0.4\textwidth, clip, trim= 0  -20 20 10]{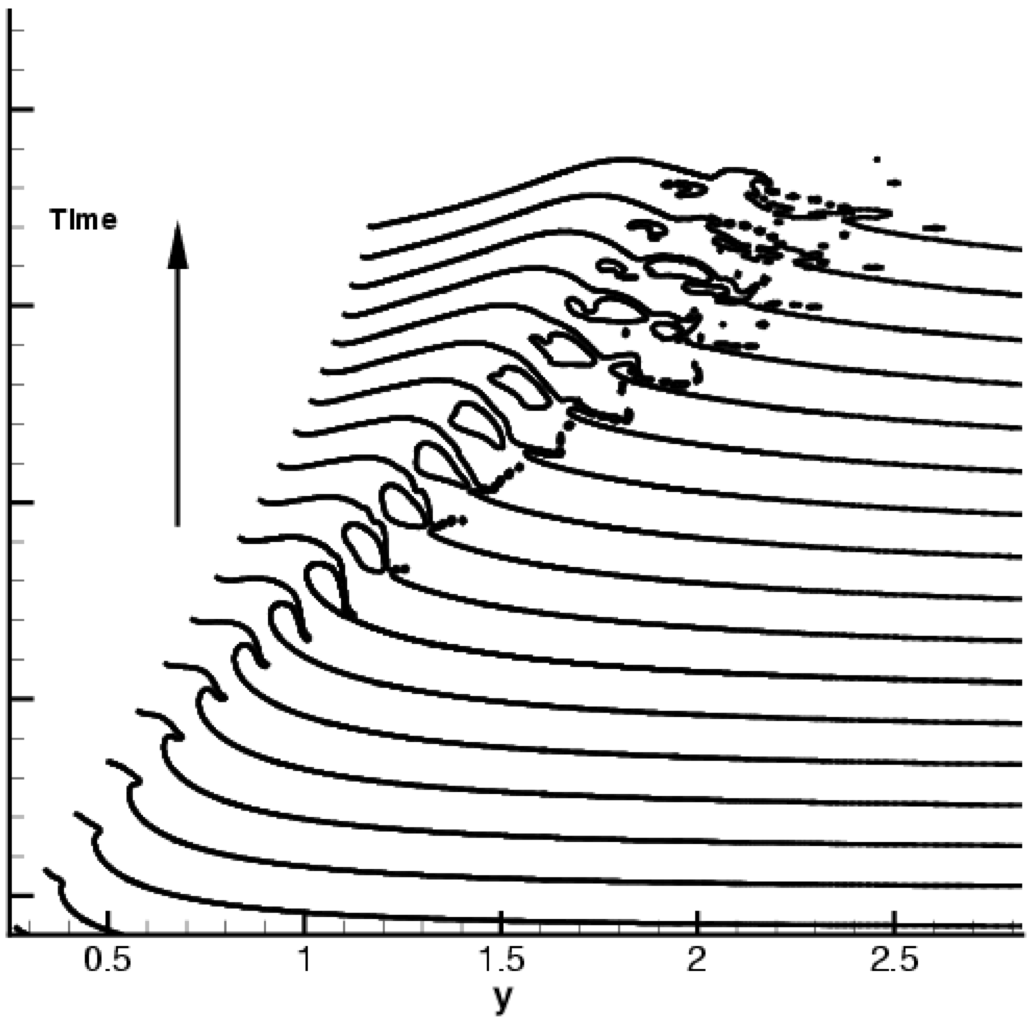}}
	\hspace{5mm}
    \subfigure[Wave crest height]{
    	\label{fig: tdptcrest b}
    	\includegraphics[width=0.4\textwidth, clip, trim= 10 10 20 10]{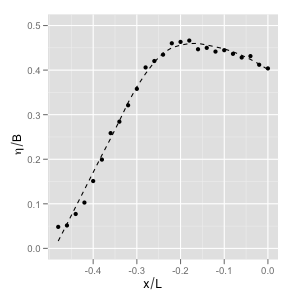}}
    \caption{2D+T results for the high speed case $F_{2D}=0.262$. (a) Waterfall plot generated by overlaying the predicted free-surface elevations at different times, shifted vertically for ease of viewing. (b) The extracted wave crest heights, with symbols for the raw data and the dashed line for the smoothed IM prediction used in $\p$.}
    \label{fig: tdptcrest}
\end{figure}

The 2D+T $\Intr$ predictions are obtained using the same Cartesian-grid method and over the same $F_{2D}$ values as in the 3D tests. While essentially the same code, the 2D+T predictions are complete in around a minute each on a desktop, as opposed to the 3D runs which take $O(10^4)$ CPU hours at a high performance computing facility. The result for the high speed case is shown in Fig.~\ref{fig: tdptcrest a}. The 2D+T model predicts a plunging breaking wave for the high speed test, the same as the 3D CFD prediction. To quantify the comparison, we extract the location of the wave crest as shown in Fig.~\ref{fig: tdptcrest b}. This the extraction introduces noise into the 2D+T results. As noted in the introduction, it is important to remove this noise with a low-pass filter to avoid the noise propagating up to the PBLM predictions. The final IM prediction is shown (as dashed lines) in Fig.~\ref{fig: tdptcrest b} and Fig.~\ref{fig: crest}. Note that the 2D+T prediction has no dependence on $B/L$ (the curves are the same for each column of Fig.~\ref{fig: crest}) and are only accurate for the slender high speed ship, as summarized in Tab.~\ref{tab: crest}.

\begin{table}
  \centering
  \begin{tabular}{rccc}
 & 2D+T & GLM & PBLM \\
$B/L = 0.03$ & 0.043 & 0.106 & 0.031 \\
$B/L = 0.06$ & 0.081 & 0.098 & 0.055 \\
$B/L = 0.12$ & 0.191 & 0.302 & 0.112 \\
  \end{tabular}
  \caption{RMSE, scaled by the maximum response of $\eta/(B F_{2D}) = 2.11$, on the test set for the breaking wave crest elevation predictions (for the high speed $F_{2D}=0.191$ case).  Despite higher error in the 2D+T IM  at greater $B/L$, PBLM outperforms GLM on the test data.}
  \label{tab: crest}
\end{table}

For the PBLM, we construct $\p$ using Eq.~\ref{eq: phase basis} with the 2D+T IM, incorporating the basis into the GLM using the weighted model Eq.~\ref{eq: weighted}, and retrained on the training data. The resulting PBLM predictions are plotted in Fig.~\ref{fig: crest} and summarized for different $B/L$ in Tab.~\ref{tab: crest}. The results show that the trained PBLM makes excellent quantitative predictions for $B/L=0.03$ where the 2D+T model gives a good estimate of the waves at all speeds. As the ship thickness increases so does the error in the 2D+T predictions, but the 2D+T IM still provides information on the dependence of the bow waves on Froude number. Complementing this information with the high and low Froude number data at each $B/L$ enables the PBLM to reduce the GLM errors by more than 60\% on the thickest hull. Thus, using PBLM with the fast 2D+T IM, we are able to generalize relatively few 3D data points across the Froude number range to generate reliable predictions that may be used, in this case, to evaluate the effect of ship geometry and speed on nonlinear breaking bow waves, at low computational cost. 

\section{Discussion and Conclusions}

We present a general method for constructing physics-based learning models (PBLM) which incorporate a physics-based basis to increase predictive accuracy and decrease training data dependence. The method requires the use of a generic learning model (GLM), a fast physics-based intermediate model (IM), and a small set of high-quality experimental or computational training data from the physical system.
We demonstrate the effectiveness of PBLM for three different ship hydrodynamics problems. To highlight the generality and versatility of the new approach, we used different sources for the training data (from experiments and computations), different generic learning models, and different intermediate models in these applications.  In each case, PBLM obtains
greatly improved prediction accuracy and robustness over GLM, and is orders of magnitude faster than high-fidelity CFD.  

In these examples we have presented only the most basic PBLM approach, using only one source of data and a single IM in each case. PBLM is directly applicable to combinations of different data (say experiments plus CFD) and multiple intermediate models (numerical, analytic or empirical). For example, the addition of an IM which described the variation in $B/L$ for the bow wave predictions of \S\ref{ssec: tdpt} would enable even higher accuracy predictions across the design space. Additionally, we have only presented one method of generating the physical basis $\p$. Techniques to establish a complete orthogonal basis, such as PCA, and use this in a PBLM framework are currently in development. 

Increased performance demands on naval and marine platforms have led to modern designs that require very large numbers of design evaluations.  This requirement is met by  tank and field experiments and high-fidelity simulations, which are expensive; complemented by approximate design and simulation tools, which do not capture all aspects of the problems or to sufficient accuracy.  The proposed PBLM provides a general and powerful framework which combines the strength of these separate approaches to obtain significantly more useful predictions than either in isolation.  This efficacy of PBLM increases the value of both high-fidelity expensive data and approximate fast tools and should prove useful in all stages of modern analysis and design.

\bigskip\noindent
{\bf Acknowledgements}
We would like to acknowledge Sheguang Zhang, Claudio Cairoli, and Thomas Fu for help supplying the intermediate and experimental sources. This research was supported financially by grants from the Office of Naval Research and the computational resources were provided through a Challenge Project grant from the DoD High Performance Computing Modernization Office.

\bibliographystyle{plainnat}
\bibliography{pblm}
\end{document}